\documentclass[graybox]{svmult}

\usepackage{mathptmx}       
\usepackage{helvet}         
\usepackage{courier}        
\usepackage{type1cm}        
\usepackage{graphicx}        
\usepackage{multicol}        
\usepackage[bottom]{footmisc}

\begin{document}

\title*{Breaking $so(4)$ symmetry without degeneracy lift}
\author{M.\ Kirchbach, A.\ Pallares Rivera, and F. de J.\ Rosales Aldape}
\institute{M.\ Kirchbach \at Institute of Physics, Autonomous University at San Luis Potosi, Av. Manuel Nava 6, SLP 78290, Mexico \email{mariana@ifisica.uaslp.mx}
\and A.\ Pallares Rivera \at Institute of Physics, Autonomous University at San Luis Potosi, Av. Manuel Nava 6, SLP 78290, Mexico \email{pallares@ifisica.uaslp.mx}
\and F.\ de J.\ Rosales Aldape \at Institute of Physics, Autonomous University at San Luis Potosi, Av. Manuel Nava 6, SLP 78290, Mexico \email{r-felipedejesus@yahoo.com.mx}}
%
%
\maketitle

\abstract{We argue that in the quantum motion of a scalar particle of mass $m$  on $S^3_R$,  perturbed by the trigonometric Scarf potential (Scarf I) 
with one internal quantized dimensionless parameter, $\ell$, the  3D orbital angular momentum,
and another, an external scale introducing continuous parameter, $B$,
a loss of  the geometric hyper-spherical $so(4)$ symmetry of the free motion
can occur that leaves intact the unperturbed ${\mathcal N}^2$-fold degeneracy patterns, with 
${\mathcal N}=(\ell +n+1)$ and $n$ denoting the nodes of the wave function.
Our point is that although the number of degenerate states for any ${\mathcal N}$  matches dimensionality of an
irreducible $so(4)$ representation space, the corresponding set of wave functions do not transform irreducibly under any $so(4)$.    
Indeed, in expanding the Scarf I wave functions in the basis of properly identified $so(4)$  representation functions, 
we find power series in the perturbation parameter, $B$, 
where 4D angular momenta  $K\in [\ell , {\mathcal N}-1]$ contribute 
up to the order $\left(\frac{2mR^2B}{\hbar^2}\right)^{{\mathcal N}-1-K}$.
In this fashion, we work out an explicit example on a symmetry breakdown by external scales  that retains the degeneracy.
The scheme extends to $so(d+2)$ for any $d$.}

\section{Introduction}
\label{sec:1}
The theory of Lie algebras provides, in terms of its invariants, a power tool for the description of observed 
constants of motion both in free and interacting systems and enables in this manner uncovering of  universal physical laws.
In spectral problems, symmetry as a rule is signaled by energy values degenerate with respect to certain sets of quantum numbers,
an indication that a Lie algebra might exist whose irreducible representations have dimensionalities  
that match the number of states in the levels. In this fashion, a relationship between symmetry and degeneracy can be established. 
Any ${ N}$-fold degenerate system is $gl({ N}, R)$ symmetric in so far as by virtue of Sturm-Liouvill's theory of differential equations,
any linear superposition of solutions characterized by a common eigenvalue is again a solution to the same eigenvalue.
The case of our interest here is the one in which the degeneracy patterns can be mapped on the irreducible representations of a Lie algebra
distinct from $gl({ N}, R)$. Popular examples are the spectra of the Harmonic-Oscillator--, and the Coulomb problems, whose Hamilton operators
can be cast in their turn as $su(3)$, and $so(4)$ invariants, respectively. Especially in the latter case, the ${\mathcal N}^2$-fold degeneracies  of the
states in a level (${\mathcal N}$ being the principal quantum number, ${\mathcal N}=n+\ell+1\in [1,\infty)$,  and $\ell$ and $n$ denoting the orbital angular momentum value,
and the number of nodes, respectively) has been explained in terms of $so(4)$ irreducible representations of dimensionalities, ${\mathcal N}^2$.
It has been realized  already in the early days of quantum mechanics  that a Hamiltonian with Coulombic interaction 
can be cast in the form of a Casimir invariant of the isometry algebra $so(4)$ of the three-dimensional (3D) sphere, $S^3_R$ with $R$ 
being the hyper-radius \cite{Fock}.
This example shows that a relationship between symmetry and degeneracy can be at the very root of spectroscopic studies, a reason for which
it is important to understand as to what extent  Lie-algebraic degeneracy patterns are at par
with the correct transformation properties of the wave functions under the algebra in question.
Our point is that degeneracy alone is not sufficient to claim a particular Lie algebraic symmetry of the Hamiltonian.
On the example of the quantum motion of a scalar particle on $S^3_R$,  perturbed by the trigonometric Scarf potential (Scarf I),
we show that the perturbation completely retains the $so(4)$  degeneracies of the free motion without that the ``perturbed'' wave functions  
would behave as eigenfunctions of an $so(4)$ Casimir operator.
 
The contribution is structured as follows. In the next section we study the $so(4)$ symmetry properties of the 
hyper-geometric differential equation for the Gegenbauer polynomials, $ {\mathcal G}_n^\lambda(x)$,
for $\lambda =(\ell +1)$ with 
$\ell$ non-negative integer.
{}First we  observe that in subjecting the eigenvalue problem of the canonical $so(4)$ Casimir operator
to a similarity transformation  by $(1-\sin^2 \chi)^{\frac{\lambda}{2} -\frac{1}{4}}$, the 
square-root of the weight function of the Gegenbauer polynomials, and setting
$x=\sin \chi$, with $\chi$ standing for the second polar angle in $E_4$,  amounts to the
Gegenbauer equation, thus making  the $so(4)$ symmetry of the latter manifest. 
As long as free quantum motion on $S^3_R$ can be cast as the eigenvalue problem of the Casimir operator of the transformed $so(4)$,  
whose  wave functions are the Gegenbauer polynomials,
$so(4)$ has been proved to be the relevant  symmetry both of the spectrum and the wave functions.
This contrasts the case of the Jacobi polynomials, $P_n^{\alpha\beta}(x)$, considered in section III for  
the following parameter values, $\alpha_\ell =\ell +\frac{1}{2}-b$, and $\beta_\ell= \ell +\frac{1}{2}+b$, 
which present themselves as linear combinations of Gegenbauer polynomials of equal $\lambda=(\ell+1)$ parameters but  different degrees, $n$, 
and do not behave as $so(4)$ representation functions. Nonetheless, because of the above specific choice of the parameters,
the Jacobi polynomial equation can be transformed to a motion on $S^3_R$ perturbed
by the trigonometric Scarf potential, whose spectrum carries by chance same $so(4)$ degeneracy patterns as the free motion, without
that this symmetry is shared by the wave functions. 
In this manner, we explicitly  work out  an example of breaking $so(4)$ by a perturbation without degeneracy lift.
 The paper closes with brief conclusions.

\section{The Gegenbauer polynomial equation as  eigenvalue problem of an $so(4)$ Casimir operator}
\label{sec:2}
The Gegenbauer polynomial equation \cite{Abr} for the special choice of the parameter, $\lambda=\ell +1$, with $\ell$ non-negative integer,
is given by 
\begin{equation}
\left( 1-x^2\right)\frac{{\mathrm d}^2{\mathcal G}_n^{\ell +1} (x)}{{\mathrm d}x^2}
-\left( 2\ell +3\right)x\frac{{\mathrm d}{\mathcal G}_n^{\ell +1} (x)}{{\mathrm d}x}
+n(n+2\ell +2){\mathcal G}_n^{\ell +1} (x)=0.
\label{GBaur}
\end{equation}
At the same time, the eigenvalue problem of the well known Casimir operator, ${\mathcal K}^2$,  of the $so(4)$ isometry algebra of the
three dimensional (3D) \underline{unit} sphere, to be denoted by $S^3$, reads
\begin{eqnarray}
\left[ {\mathcal K}^2 -K(K+2)\right]Y_{K\ell m}(\chi, \theta, \varphi)=0,\, \,  
{\mathcal K}^2=\frac{(-1)}{\cos^2\chi}\frac{\partial}{\partial \chi}\cos^2\chi \frac{\partial}{\partial \chi}  +
\frac{{\mathbf L}^2(\theta,\varphi)}{\cos^2\chi},&&  \nonumber\\
Y_{K\ell m}(\chi, \theta,\varphi)= \cos^{\ell }\chi {\mathcal G}_{n=K-\ell}^{\ell +1}(\sin \chi)Y_{\ell m}(\theta, \varphi),
\,\, \quad K=n+\ell,&&
\nonumber\\                  
{\mathbf L}^2(\theta,\varphi)Y_{\ell m}(\theta,\varphi)=\ell(\ell +1)Y_{lm}(\theta,\varphi).\,\,\qquad\qquad &&
\label{CasS3}
\end{eqnarray} 
Here, ${\mathbf L}(\theta,\varphi)$ is the 3D angular momentum operator, $K$, $\ell$ and $m$ are in turn the 4D-, 3D, and 2D 
angular momentum values, $Y_{K\ell m}(\chi,\theta, \varphi)$ are the 4D spherical harmonics, with $\chi\in \left[-\frac{\pi}{2},+\frac{\pi}{2}\right]$, and $\theta\in \left[ 0,\pi \right] $  
standing for the two polar angles parametrizing $S^3$, and $\varphi\in \left[ 0,2\pi \right]$ denoting the ordinary azimuthal angle.  
In the so called quasi-radial variable \cite{Kalnin}, $\chi$, the equation ~(\ref{CasS3}) reduces to
 
\begin{eqnarray}
\left[-\frac{1}{\cos^2\chi}\frac{\partial}{\partial \chi}\cos^2\chi \frac{\partial}{\partial \chi} +
\frac{\ell (\ell +1)}{\cos^2\chi}-K(K+2)\right]\cos^{\ell}\chi {\mathcal G}_{K-\ell}^{\ell +1}(\sin\chi)&=&0,
\label{GeB1}
\end{eqnarray}
and it is straightforward to check that (\ref{GeB1}) is equivalent to
\begin{eqnarray}
\left[ {\widetilde {\mathcal K}}^2-(n+l)(n+l+2))\right]{\mathcal G}_{K-\ell}^{\ell +1}(\sin\chi)&=&0,\nonumber\\
\mbox{ with} \qquad {\widetilde {\mathcal K}}^2=\cos^{-\ell}\chi {\mathcal K}^2\cos^{\ell} \chi,&&
\label{GeB3}
\end{eqnarray}
because of 
\begin{eqnarray}
{\widetilde {\mathcal K}}^2=\cos^{-\ell}\chi {\mathcal K}^2\cos^\ell \chi &=&
-\frac{{\mathrm d}^2}{{\mathrm d}\chi^2}
+\left( 2\ell +2\right)\tan\chi \frac{{\mathrm d}}{{\mathrm d}\chi}
+\ell (\ell +2).
\label{GeB2}
\end{eqnarray}
The $\cos^\ell\chi$ function relates to the square-root of the weight function, $\omega^\lambda (x)$, of the Gegenbauer polynomials, ${\mathcal G}_n^\lambda (x)$, as,
\begin{eqnarray}
\omega ^\lambda (x) =(1-x^2)^{\lambda-\frac{1}{2} },  &\quad&  x=\sin\chi, \quad  \lambda =(\ell +1),\nonumber\\ 
\cos ^\ell \chi &=&\sqrt{
\frac{\omega ^{\ell +1}(\sin\chi)}
{\frac{
{\mathrm d}x}{{\mathrm d}\chi}
}}.
\label{sqrtwafu}
\end{eqnarray}
 Therefore, upon changing variable in (\ref{GeB2}) to $x=\sin\chi$, and back-substituting in (\ref{GeB1}),
one obtains the claimed equality between the  $so(4)$ Null operator,
\begin{equation}
\left[{\widetilde {\mathcal K}}^2-K(K+2)\right], \quad K=n+\ell, 
\label{Nulloprt}
\end{equation}
and  the Gegenbauer polynomial equation as,
\begin{eqnarray}
\left[ {\widetilde {\mathcal K}}^2-(n+l)(n+l+2)\right]{\mathcal G}_{K-\ell}^{\ell +1}(x)
&=&
\left( 1-x^2\right)\frac{{\mathrm d}^2{\mathcal G}_{K-\ell}^{\ell +1} (x)}{{\mathrm d}x^2}-\left( 2\ell +3 \right)
\frac{{\mathrm d}{\mathcal G}^{\ell +1}_{K-\ell} (x)}{{\mathrm d}x}\nonumber\\
&+&n(n+2\ell +2){\mathcal G}_{{K-\ell}}^{\ell +1} (x)=0.
\label{Gegenb_so4symm}
\end{eqnarray}
The latter equation means that the Gegenbauer polynomials, occasionally termed to as ultra-spherical polynomials, are
representation functions to an $so(4)$ algebra obtained from the canonical one according to (\ref{sqrtwafu})
through a similarity transformation by the square-root of their weight function and upon  accounting for a change of variable.     
An interesting connection between the latter equation and the 1D Schr\"odinger equation with the
$\sec^2\chi$ potential can be established upon substituting, 
\begin{equation}
\cos^{\ell} \chi {\mathcal G}_{K-\ell}^{\ell+1}(\sin\chi) =\frac{{\mathcal U}_n^{\ell}(\chi)}{\cos\chi}, \quad n=K-\ell.
\label{S3_1DSchr}
\end{equation}
In so doing, one finds that ${\mathcal U}_n^{\ell}(\chi)$ satisfies the  1D Schr\"odinger equation with the $\sec^2\chi$ potential 
according to,  
\begin{equation} 
\left[-\frac{{\mathrm d}^2}{{\mathrm d}\chi^2}
+\frac{\ell (\ell +1)}{\cos^2\chi}\right] {\mathcal U}_n^{\ell}(\chi)= (K+1)^2 {\mathcal U}_n^{\ell}(\chi), 
\label{secantaq}
\end{equation}
whose spectrum is characterized by $(K+1)^2$-fold degeneracy of the levels, just as the H atom, due to
$\sum_{\ell=0}^{\ell=K}(2\ell +1)=(K+1)^2$. 
Therefore, the $so(4)$ symmetry of the Gegenbauer polynomials shows up as $so(4)$ degeneracy patterns in the spectrum of the 
corresponding 1D Schr\"odinger equation with the
$\sec^2\chi$ interaction. 
More general, there are  several two-parameter potentials, $v(z;\alpha,\beta)$ for which the Schr\"odinger equation,
\begin{equation}
\left[-\frac{{\mathrm d}^2}{{\mathrm d}z^2} +v(z;\alpha, \beta))\right] R^{\alpha\beta}_n(z)= \epsilon R^{\alpha\beta}_n(z), 
\label{gen_Schr}
\end{equation}
can be exactly solved by reducing it
to a hyper-geometric differential equation by
means of a point-canonical transformation of the type \cite{Levai},
\begin{equation}
R_n^{\alpha\beta}(z)=R_n^{\alpha\beta}\left( z=f(x)\right):
\stackrel{\mbox{\footnotesize def}}{=}g^{\alpha\beta}_n(x)=\sqrt{\omega^{\alpha\beta}(x)}J_n^{\alpha \beta}(x)
\frac{1}{\sqrt{
\frac{{\mathrm d}f(x)}{{\mathrm d}x}}},\quad x\in [a,b],
\label{point_can}
\end{equation}
where $J_n^{\alpha\beta}(x)$ are  polynomials of degree $n$ and orthogonal with respect to their weight-function $\omega^{\alpha\beta}(x)$ according to
\begin{equation}
\int_0^\infty R^{\alpha\beta}_n(z)R^{\alpha\beta}_{n^\prime}(z){\mathrm d}z=
\int_a^bg_n^{\alpha\beta}(x)g_{n^\prime}^{\alpha\beta}(x){\mathrm d}f(x)=
\int_a^b\omega ^{\alpha\beta}(x)J_n^{\alpha\beta}(x)J_{n^\prime}^{\alpha\beta}(x){\mathrm d}x.
\label{J_ort}
\end{equation}
And vice verse, any hyper-geometric differential equation  can be brought back to an 
1D Schr\"odinger equation in (\ref{gen_Schr}) by inverting the transformation in (\ref{point_can}).

\noindent
The above procedure establishes an interesting link between the symmetry properties of orthogonal polynomials
and the degeneracies in the corresponding  potential spectra.
In the next subsection we shall see that a Lie algebraic degeneracy in the Schr\"odinger spectrum can appear by chance
and without it being shared by the polynomial equation.

\section{A Jacobi polynomial equation as eigenvalue problem of a ``frustrated'' $so(4)$ Casimir operator}
The hyper-geometric differential equation solved by the Jacobi polynomial reads \cite{Abr},
\begin{equation}
\left( 1-x^2\right)\frac{{\mathrm d}^2{ P}_n^{\alpha \beta} (x)}{{\mathrm d}x^2}
+\left[(\beta -\alpha)-\left( \alpha +\beta +2 \right)x\right]\frac{{\mathrm d}{P}_n^{\alpha \beta} (x)}{{\mathrm d}x}
+n(n+\alpha  +\beta +1){P}_n^{\alpha \beta} (x)=0,
\label{Jac}
\end{equation}
and acquires a shape pretty close to (\ref{GBaur}) for the following special choice  of the parameters,
\begin{equation}
\alpha_\ell   =\ell -b +\frac{1}{2}, \quad \beta_\ell  =\ell +n +\frac{1}{2},
\label{Jac_parms}
\end{equation}
namely,
\begin{eqnarray}
{\Big(}{\widetilde {\mathcal K}^2}-(n+\ell)(n+\ell +2) +2b \frac{{\mathrm d}}{{\mathrm d}x}{\Big)}
{P}_n^{\ell -b+\frac{1}{2}, \ell +b+\frac{1}{2}} (x)&&\nonumber\\
=\left( 1-x^2\right)\frac{{\mathrm d}^2{ P}_n^{\ell -b+\frac{1}{2}, \ell +b \frac{1}{2} } (x)}{{\mathrm d}x^2}
+\left[2b-\left( 2\ell +3 \right)x\right]\frac{{\mathrm d}{P}_n^{\ell -b+\frac{1}{2}, \ell +b +\frac{1}{2}} (x)}{{\mathrm d}x}&&\nonumber\\
+n(n+2\ell +2){P}_n^{\ell -b +\frac{1}{2}, \ell +b+\frac{1}{2}} (x)=0.&&
\label{Jac_spec}
\end{eqnarray}
The latter relation reveals the Jacobi equation as the $so(4)$ Null-operator in (\ref{Nulloprt}), ``frustrated'' by the
gradient term $\left[-2b\frac{{\mathrm d}}{{\mathrm d}x}\right]$. In consequence, the Jacobi polynomials do not behave as $so(4)$
representation functions. This  is best illustrated through the finite series decomposition of a Jacobi polynomial of degree $n$ 
in Gegenbauer polynomials of degrees running from $0$ to $n$, shown in Table I.
In recalling that the degrees of the  Gegenbauer polynomials under considerations
express in terms of the 4D angular momentum values, $K$, as $n=(K-\ell)$, the decompositions present themselves as
mixtures of $so(4)$ representation functions of different 4D angular momentum values, 
$K\in \left[\ell, \ell +n\right]$.
   
\noindent
Despite the absence of $so(4)$ symmetry of the Jacobi polynomials,
a curiosity occurs insofar as the associated 1D Schr\"odinger equation (in units of $\hbar^2/(2mR^2)$),
\begin{eqnarray}
\left[- \frac{{\mathrm d}^2}{{\mathrm d}\chi^2}+ v_{\mbox{\footnotesize ScI}}\left(\chi ;\alpha_\ell,\beta_\ell \right)\right]\,
R_n^{\ell-b+\frac{1}{2},\ell +b+\frac{1}{2}} \left( \chi \right)=
\epsilon \, R_n^{\ell-b+\frac{1}{2},\ell +b+\frac{1}{2}} \left( \chi \right),&&\label{Schr1}\\
 v_{\textup{ScI}}\left(\chi;\alpha_\ell,\beta_\ell \right)=
\frac{b^2+ \ell(\ell+1)}{\cos^2 \chi } -\frac{b(2\ell+1)\tan \chi }{\cos \chi },\,\,
b=\frac{B(2mR^2)}{\hbar^2},&&
\label{Schr_prmts}\\
R_n^{\ell-b+\frac{1}{2},\ell +b+\frac{1}{2}} \left( \chi \right)=e^{-b\tanh^{-1}\sin\chi}
\cos^{\ell +1}\chi P_n^{\ell-b+\frac{1}{2}, \ell+b+\frac{1}{2}}\left(\sin \chi  \right),&&
\label{Schr3}\\
{\mathcal N}=n+\ell +1, \quad  \epsilon = {\mathcal N}^2, \quad\epsilon=\frac{E(2mR^2)}{\hbar^2},
[E], [B]=\mbox{MeV},&&
\label{Scarfs_pot}
\end{eqnarray}
reduced to the hyper-geometric differential equation  along the line of the above eqs.~(\ref{J_ort})--(\ref{point_can})
for $\chi =f(x)=\sin^{-1}x$, and
\begin{equation}
\omega^{\ell -b+\frac{1}{2}, \ell +b +\frac{1}{2}}(x)=
e^{-b\tanh^{-1}x} (1-x^2)^{\ell +\frac{1}{2}},
\label{Jac_weightf}
\end{equation}
exhibits same degeneracy patterns as the fully $so(4)$ symmetric problem in (\ref{secantaq}) and the underlying (\ref{Gegenb_so4symm}).
In (\ref{Schr_prmts}), the $v_{\mbox{\footnotesize ScI}}\left(\chi;\alpha_\ell,\beta_\ell \right)$ potential is known under the name of the trigonometric Scarf potential,
abbreviated, Scarf I (\cite{Levai} and references therein). Under the substitution, 
\begin{equation}
R_n^{\ell-b+\frac{1}{2},\ell +b+\frac{1}{2}} \left( \chi \right)=\frac{U_n^{\ell-b+\frac{1}{2},\ell +b+\frac{1}{2}} \left( \chi \right)}{\cos\chi},
\label{flat_hyp}
\end{equation}
the equation (\ref{Schr_prmts}) is transformed to motion on $S^3$ perturbed by Scarf I.
The expansions in Table I apply equally well to the wave functions $U_n^{\ell-b+\frac{1}{2},\ell +b+\frac{1}{2}} \left( \chi \right)$
 which can not transform
as $so(4)$ representation functions despite the $so(4)$ degeneracy patterns in the spectrum.
In this fashion, we worked out an example that a Lie algebraic symmetry in a spectral problem does not necessarily
imply same symmetry of the Hamiltonian. The Figure 1 is illustrative of  this type of  $so(4)$ breaking. 
\begin{table}
\caption{Decompositions of  some of the Jacobi polynomials $P_n^{\ell -b+\frac{1}{2},\ell +b +\frac{1}{2}}(\sin\chi)$ in  (\ref{Jac_parms})
for fixed $\ell$  in $so(4)$ representation functions of 4D angular momenta 
$K\in \left[ \ell \, , \kappa \right] $ with $\kappa =n+\ell={\mathcal N}-1$.
The $K$ labeled Gegenbauer polynomials contribute to the order ${\mathcal O}\left( b^{\kappa -K}\right) $ to the expansion and 
give the  order to which the $so(4)$ symmetry fades away, with $b$ defined in (\ref{Jac_parms}) and  (\ref{Schr_prmts}).}
\label{TabJac}  
\begin{tabular}{p{1.5cm}p{2.4cm}p{3.2cm}p{0.5cm}p{4.1cm}}
\hline
 $\ell$ &  K&
$P_{n=\kappa-{ \ell}}^{(\alpha_{\ell},\beta_{\ell})}(\sin \chi)$
&=&$\sum_{{K} = {\ell}}^{\kappa }c_{K \kappa }(b)\,\mathcal{G}_{K -{\ell}}^{{\ell}+1}(\sin \chi)$ \\
\hline   \hline
&&&& \\
 $\kappa$  & K=$\kappa$ & $P_{0}^{(\alpha_k,\beta_k)}(\sin \chi)$ &=& $\mathcal{G}_{0}^{k+1}(\sin \chi)$ \\
&&&& \\
 $\kappa-1 $& K$\in\left[\kappa -1, \kappa\right]$ & $P_{1}^{(\alpha_{k-1},\beta_{k-1})}(\sin \chi)$
&=&
$\frac{(2k+1)}{4k}\,\mathcal{G}_{1}^{k}(\sin \chi)-b\,\mathcal{G}_{0}^{k}(\sin \chi)$\\
&&&& \\
 $\kappa-2$ & K $\in \left[\kappa -2,\kappa\right]$ &$ P_{2}^{(\alpha_{k-2},\beta_{k-2})}(\sin \chi)$
&=&
${1\over 8}\frac{(2k+1)}{(k-1)}\,\mathcal{G}_{2}^{k-1}(\sin \chi)-{b\over 2} \frac{k}{(k-1)}\,\mathcal{G}_{1}^{k-1}(\cot \chi)
+ \frac{b^2}{2}\,\mathcal{G}_{0}^{k-1}(\sin \chi)$ \\
&&&& \\
$\kappa-3$ & K$\in \left[ \kappa-3, \kappa\right]$ &  $P_{3}^{(\alpha_{k-3},\beta_{k-3})}(\sin \chi)$
&=&
${1\over 32} \frac{(4\kappa^2-1)}{(\kappa^2-3\kappa + 2)}\,\mathcal{G}_{3}^{(k-2)}(\sin \chi) 
 - {b \over 8} \frac{(2\kappa^2-\kappa)}{(\kappa^2-3\kappa +2)}\,\mathcal{G}_{2}^{(k-2)}(\sin \chi) 
      +{b^2 \over 8}\frac{(2k-1)}{(k-2)}\,\mathcal{G}_{1}^{(k-2)}(\sin \chi)$ \\[2mm]
&&&& \,$-{b \over 24 }\frac{(4b^2 k -4b^2 + 2k +1)}{(k-1)}\,\mathcal{G}_{0}^{(k-2)}(\sin \chi)$ \\
&&&& \\
\hline
\hline
\end{tabular}

\end{table}

%
\begin{figure}[b]
\sidecaption
\includegraphics[scale=.4]{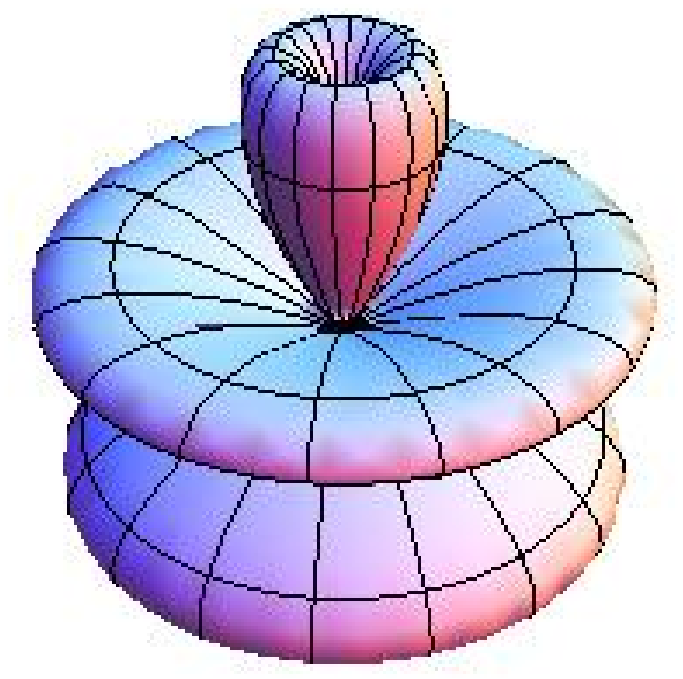}
\includegraphics[scale=.25]{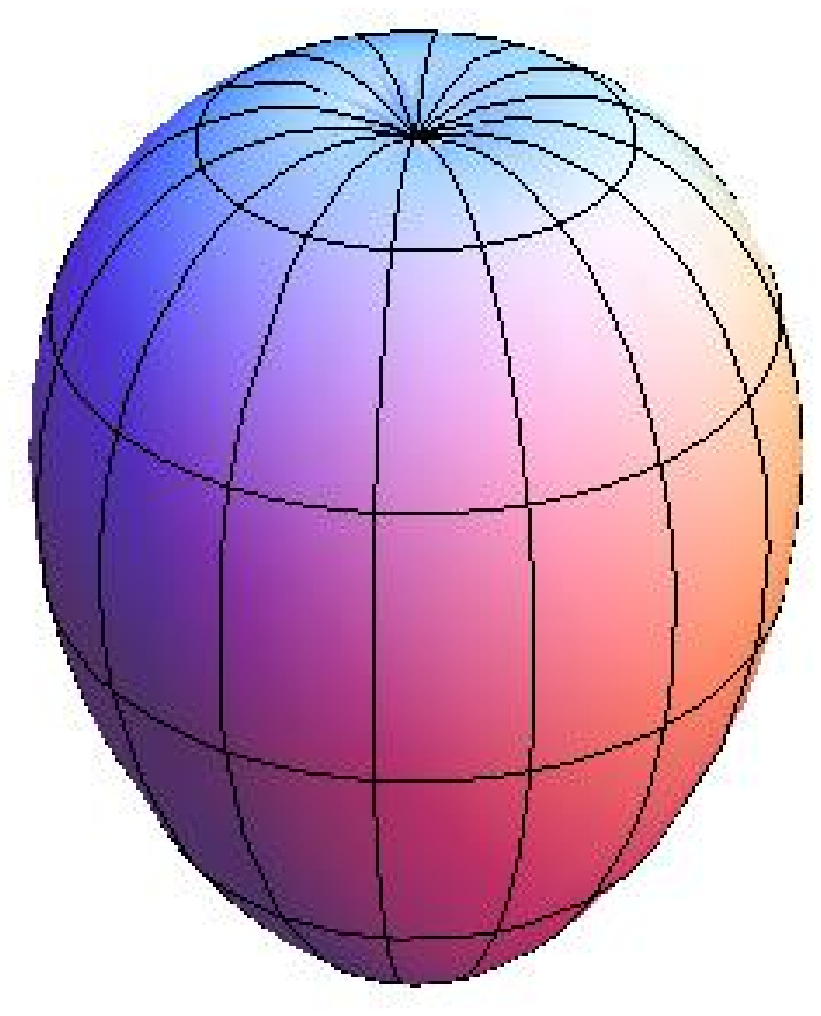}
%
%
\caption{The breaking of the $so(4)$ symmetry of the free motion of a scalar particle on $S^3$ in (\ref{GeB1})-(\ref{GeB2}) and (\ref{secantaq}),
through the external scale $B=\frac{\hbar ^2 b}{2mR^2}$, due to a perturbation by the trigonometric Scarf potential (\ref{Schr1})--(\ref{Scarfs_pot}). 
The wave function $U_3^{\frac{3}{2}-b,\frac{3}{2}+b }(\chi)$ in (\ref{flat_hyp}) (right) in comparison to
its counterpart, $\cos^{\ell}\chi {\mathcal G}_3^{2}(\sin \chi)$ in (\ref{GeB3}) (left) describing the unperturbed  $so(4)$ symmetric 
motion. 
These functions describe equal energies in the respective potential problems in eqs.~(\ref{secantaq}) and (\ref{Scarfs_pot}).}
\label{fig:1}       
\end{figure}

\begin{figure}[h]
\sidecaption[t]
\centering{
\includegraphics[scale=.45]{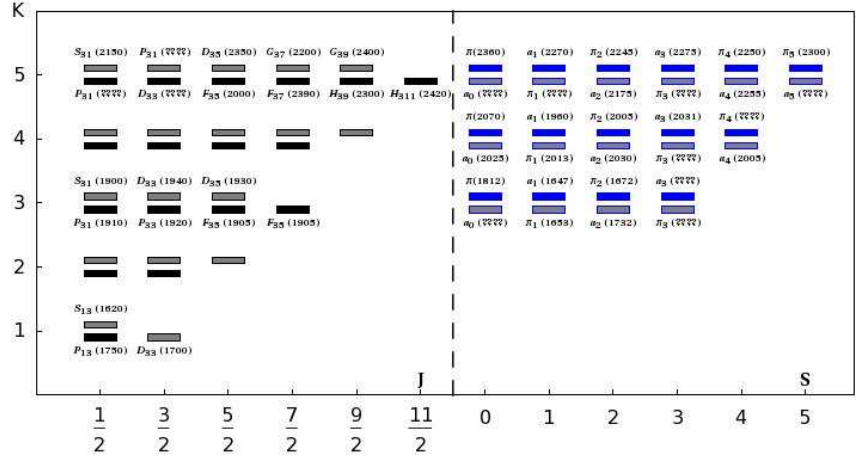}}
%
%
\caption{Hydrogen like (conformal type) degeneracy in the reported spectra of the excited 
$L_{3(2J)}$, i.e.
$\Delta$ baryons (left)
 and the high-lying
light flavored mesons (right) (for details on the  notations and  more references see \cite{Kirchbach2010},\cite{PART}).  Full and shadowed bricks denote degenerate hadron 
states of opposite parities. The numbers  inside of the parenthesis give the masses (in MeV) while
the question marks denote ``missing'' states. The meson sector is close to
parity doubled,  a possible hint on chiral symmetry 
restoration from the spontaneously broken Nambu-Goldstone-- to the manifest Wigner-Wyle mode. 
Remarkable, the pronounced supersymmetric baryon-meson degeneracy.}
\label{fig:2}       
\end{figure}

\begin{figure}[h]
\begin{center}
 \includegraphics[scale=.4]{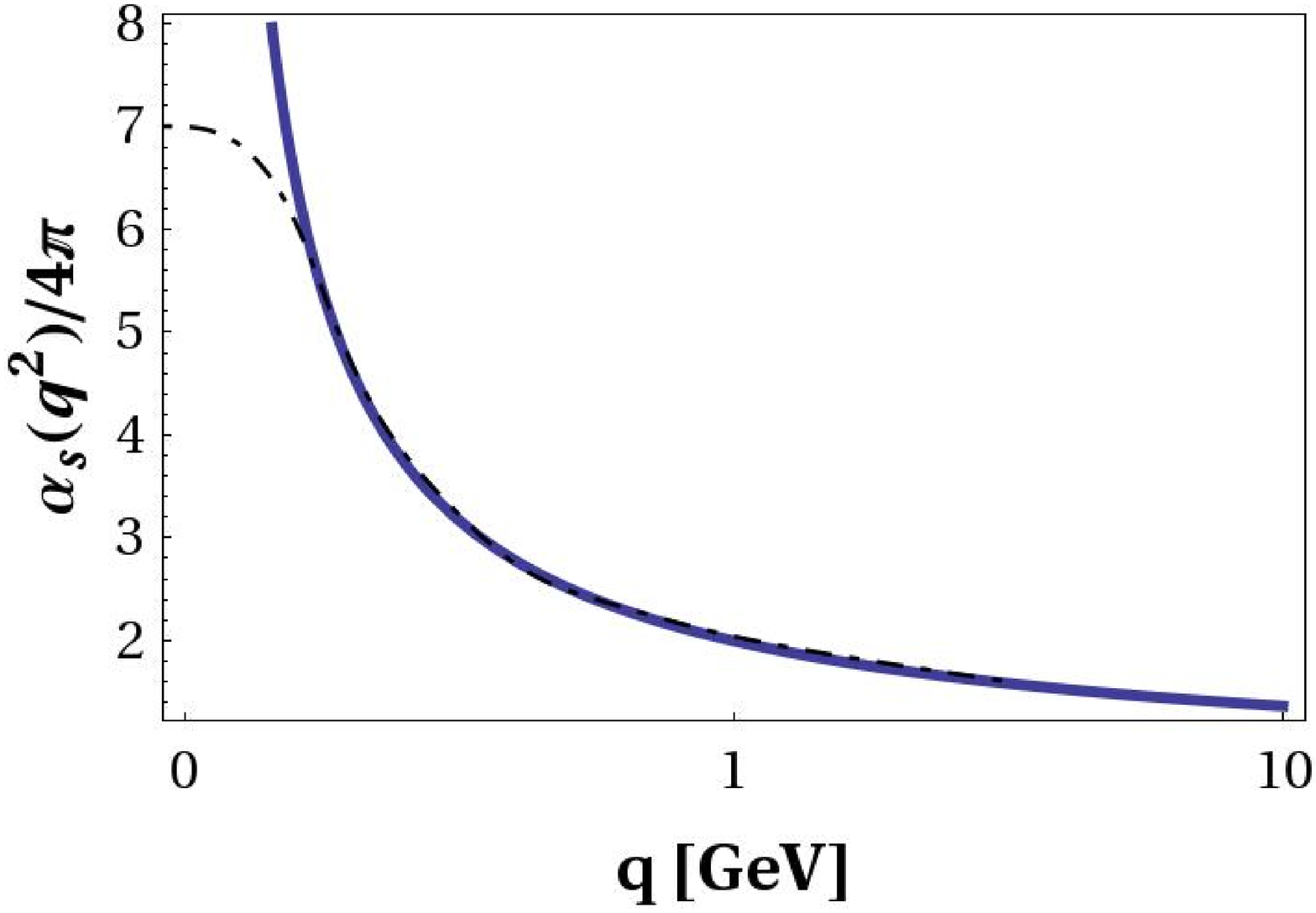}
\end{center}

%
%
\caption{Schematic presentation of the walking (dashed line) of the strong coupling constant in the infrared according to \cite{Deur}.}
\label{fig:3}       
\end{figure}

\section{Conclusions}
In this work we constructed an explicit example for the possibility to remove a Lie algebraic symmetry of a 
Hamiltonian by perturbation and without lifting the unperturbed degeneracy patterns in the spectrum.
The clue of this observation is that Lie algebraic degeneracy patterns  can throughout be tolerant towards external scales, such as masses, 
temperatures, lengths etc. Such a type  of $so(4)$ symmetry lift
could reconcile the experimentally detected conformal symmetry patterns in the spectra of the high-lying light flavored hadrons, 
both baryons and mesons, with the conformal symmetry removal through the dilaton mass. The relevance of the conformal symmetry for QCD
is predicted by the AdS$_5$/CFT$_4$ duality and is compatible with spectroscopic data on the light-flavored hadron spectra (see Fig.~2) due to
the walking of the strong coupling constant in the infrared towards a fixed value \cite{Deur},  sketched in Fig.~3. 
The relevance of the hyper-spherical geometry in conformal field theories is derived from the possibility of mapping a flat space-time 
QFT on Einstein's closed universe, $ {\mathcal R}^1\otimes S^3_R$, whose isometry algebra is the covering  of the conformal one, a result due to \cite{LuMack}.
The so called compactified Minkowski space time, in being of finite 3D volume, provides a natural scenario for the QCD confinement phenomenon
\cite{Witten} and the inverse of the $S^3_R$ radius provides a natural scale that can be interpreted as the temperature \cite{Tommy}.

\begin{acknowledgement}
One of us (M.K.) thanks the organizers of the ``Lie Theory and Its Applications In Physics'' 2013 conference in Varna for their
hospitality and efforts. We are indebted to Dr. Jose Antonio Vallejo for helpful comments and Jose Limon Castillo for
assistance in computer matters.
\end{acknowledgement}
\end{document}